\documentclass{jfm}
\usepackage{graphicx}
\usepackage{epstopdf, epsfig}
\usepackage{amsmath,mleftright}
\usepackage{xparse}
\usepackage{caption}
\usepackage[normalem]{ulem}
\usepackage{natbib}
\usepackage{mathtools} %for rcases
\usepackage{comment} % to comment out large blocks
\usepackage{hyperref}

\newcommand{\what}{\widehat}

\usepackage{xcolor}
\definecolor{light-gray}{gray}{0.5}
\definecolor{blue}{rgb}{0.0,0.0,1.0}
\definecolor{green}{rgb}{0.0,0.5,0.0}
\definecolor{red}{rgb}{1.0,0.0,0.0}
\definecolor{cyan}{rgb}{0.0,0.75,0.75}
\definecolor{magenta}{rgb}{0.75,0.0,0.75}
\definecolor{yellow}{rgb}{0.75,0.75,0.0}
\definecolor{orange}{rgb}{0.9,0.3,0.0}

\newcommand{\Ra}{\mathit{Ra}}
\renewcommand{\Pr}{\mathit{Pr}}

\shorttitle{The onset of zonal modes in two-dimensional Rayleigh--B\'enard convection}
\shortauthor{P.~Winchester, P.~D.~Howell and V.~Dallas}

\title{The onset of zonal modes in two-dimensional Rayleigh--B\'enard convection}

\author{Philip Winchester\aff{1}
  \corresp{\email{winchester@maths.ox.ac.uk}},
 \and Peter~D.~Howell\aff{1}
 \corresp{\email{howell@maths.ox.ac.uk}},
   Vassilios Dallas \aff{1}
  \corresp{\email{dallas@maths.ox.ac.uk}}}

\affiliation{\aff{1}Mathematical Institute, University of Oxford, Oxford, OX2 6GG, UK
}

\begin{document}

\maketitle

\begin{abstract}
We study the stability of steady convection rolls
in 2D Rayleigh--B\'enard convection with free-slip boundaries and horizontal periodicity over twelve orders of magnitude in the Prandtl number $(10^{-6} \leq \Pr \leq 10^6)$ and five orders of magnitude in the Rayleigh number $(8\pi^4 < \Ra \leq 3 \times 10^7)$. The analysis is facilitated by partitioning our modal expansion into so-called even and odd modes. With aspect ratio $\Gamma = 2$, we observe that zonal modes (with horizontal wavenumber equal to zero) can emerge only once
the steady convection roll state consisting of even modes only becomes unstable to odd perturbations. We determine the stability boundary in the $(\Pr,\Ra)$-plane and observe remarkably intricate features corresponding to qualitative changes in the solution,
as well as three regions where the steady convection rolls lose and subsequently regain stability as the Rayleigh number is increased. We study the asymptotic limit $\Pr \to 0$ and find that the steady convection rolls become unstable almost instantaneously, eventually leading to non-linear relaxation osculations and bursts, which we can explain with a weakly non-linear analysis. In the complementary large-$\Pr$ limit, we observe that the stability boundary reaches an asymptotic value $\Ra = 2.54 \times 10^7$ and that the zonal modes at the instability switch off abruptly at a large, but finite, Prandtl number. 
\end{abstract}

\begin{keywords}
B\'enard convection, bifurcations, low-dimensional models
%[please see list of JFM keywords at \url{https://www.cambridge.org/core/services/aop-file-manager/file/605b554be44d662e83c26750/JFM-Keywords-2021.pdf}]
\end{keywords}
\section{Introduction}

Rayleigh--B\'enard convection typically begins with a steady cellular pattern (for example, rolls, squares or hexagons), and the stability of these cellular patterns has been studied for decades \citep{Busse67,Busse83,Busse84,Bolton85,Rucklidge96,Paul12}. In a 2D periodic box, the cellular pattern takes the form of convection rolls which are invariant under reflections both in the vertical plane that separates them and in the horizontal mid-plane. The vertical mirror symmetry can be broken in a pitchfork bifurcation, generating a net zonal flow in which any motion in one direction at the top will be balanced by an equal and opposite motion at the bottom \citep{Rucklidge96}. The physical mechanism behind this instability is well understood \citep{Thompson70,Busse83,Howard86}. Suppose a pair of initially symmetric rolls
tilts over, say to the right. The rising plume
will now transport rightward momentum to the top of the layer, while the descending plume will transport leftward momentum to the bottom of the layer. The resulting horizontal streaming motion causes a net zonal flow across the layer, which may be enough to sustain the original tilt of the rolls. The vertical mirror symmetry may also be broken in a Hopf bifurcation, leading to oscillations in which the direction of the zonal flow alternates \citep{Landsberg91,Proctor93, Rucklidge96}.

Large scale zonal flow in buoyancy-driven convection has been found in the atmosphere of Jupiter \citep{Heimpel05, Kong18, Kaspi18}, the Earth's oceans \citep{Max05, Richardson06, Nadiga06}, nuclear fusion devices \citep{diamondetal05, fujisawa08}, and laboratory experiments \citep{Zhang20, Read15, Krishnamurti1981}. The shearing instability which generates this net zonal flow has been studied both in the full Boussinesq equations \citep{Rucklidge96, Paul12} and in several modal truncations \citep{Howard86,Hermiz95,Horton96,Rucklidge96,Aoyahi97,Berning00}. More recently, prominent zonal flow has been found to greatly suppress convective heat transfer \citep{Gol14}
and
depend strongly on the geometry of the studied domain \citep{Lohse20,Fuentes21}.

\citet{Winchester21} showed how zonal flow can emerge at Prandtl number $\Pr=30$ in two-dimensional Rayleigh--B\'{e}nard convection through a sequence of bifurcations as Rayleigh number $\Ra$ increases. First the system undergoes a Hopf bifurcation, resulting in an oscillating zonal flow, which then grows in amplitude until the system becomes attracted to one of the two symmetric metastable shearing states. At intermediate Rayleigh numbers, the system performs apparently random-in-time transitions between these states, resulting in abrupt zonal flow reversals. As $\Ra$ increases further, the system ultimately converges to one of the shearing states, resulting in a persistent zonal flow.

In this article, we study in further detail the initial emergence of zonal modes, with horizontal wavenumber equal to zero. We analyse the stability of steady two-dimensional convective rolls over twelve orders of magnitude in the Prandtl number $(10^{-6} \leq \Pr \leq 10^6)$ and five orders of magnitude in the Rayleigh number $(8\pi^4 < \Ra \leq 3 \times 10^7)$.
We focus on the case of free-slip boundary conditions on the horizontal boundaries and periodic boundary conditions on the vertical boundaries in a rectangular domain of width-to-height aspect ratio $\Gamma =2$.

The equations governing Rayleigh--B\'enard convection and our modal decomposition are outlined in \S\ref{sec:Problem_Formulation}.
In \S\ref{sec:Linear_stability}, we describe the numerical scheme used to calculate steady states and their stability. We determine the stability boundary in the $(\Pr,\Ra)$-plane and demonstrate how remarkably intricate features on the boundary correspond to qualitative changes in the solution.
In \S\ref{sec:small} the regime of small Prandtl number is examined in more detail, by using formal asymptotic analysis to construct a system of two amplitude equations in the limit $\Pr \to 0$.
The complementary limit of $Pr \to \infty$ is studied in \S\ref{sec:large}.
Concluding remarks on the article's findings appear in \S\ref{sec:Conclusions}.

\section{Problem formulation}\label{sec:Problem_Formulation}
We consider  two-dimensional Rayleigh--B\'enard convection governed by the dimensionless equations
\begin{subequations}
\label{eq:gov}
\begin{align}
    \nabla^{2}\psi_t + \{ \psi, \nabla^{2}\psi\} & =  \Ra\Pr\theta_x + \Pr\nabla^4 \psi,\\
    \theta_t + \{ \psi, \theta\} &= \psi_x + \nabla^2 \theta,  
\end{align}
\end{subequations}
where $\psi(x,y,t)$ is the streamfunction and $\theta(x,y,t)$ is the field of temperature fluctuations from the heat-conducting temperature profile $T = 1 - y + \theta$.
Here, subscripts denote partial derivatives, and $\{f, g\} = f_x g_y - g_x f_y$ is the usual Poisson bracket. Equations \eqref{eq:gov} have been non-dimensionalised using $d$, $d^2/\kappa$ and $\Delta T$ as the relevant scales for length, time and temperature, respectively, where $d$ is the height of the fluid layer, $\kappa$ is the thermal diffusivity and $\Delta T$ is the temperature difference imposed across the layer. The two dimensionless parameters in the system \eqref{eq:gov} are the Prandtl and Rayleigh numbers
\begin{equation}
  \Pr = \frac{\nu}{\kappa}, 
  \qquad 
  \Ra = \frac{\alpha \Delta T g d^3}{\nu \kappa},
  \label{Params}    
\end{equation}
where $\nu$ is the kinematic viscosity, $\alpha$ is the thermal expansion coefficient, and $g$ is the gravitational acceleration.

The dimensionless spatial domain of our problem is $(x, y) \in [0, \Gamma] \times [0, 1]$, where $\Gamma$ is the width-to-height aspect ratio. The aspect ratio in this study is fixed at $\Gamma = 2$. At the lower ($y = 0$) and upper ($y = 1$) boundaries the temperature satisfies isothermal conditions while the velocity field satisfies no-penetration and stress-free boundary conditions, i.e.
\begin{equation}
  \theta \big |_{y=0,1} = 0, 
  \qquad 
  \psi \big |_{y=0,1} = \psi_{yy} \big |_{y=0,1} = 0,
    \label{BCs}
\end{equation}
and periodic boundary conditions are imposed in the $x$-direction.

Given the above boundary conditions, it is convenient to decompose the streamfunction into basis functions with Fourier modes in the $x$-direction and sine modes in the $y$-direction, viz.
\begin{align}
    \psi(x,y,t) &= \sum_{k_x\in\mathbb{Z}}\; \sum_{k_y\in\mathbb{Z}_{>0}} \what{ \psi}_{k_x,k_y}(t) \mathrm{e}^{2\pi \mathrm{i} k_x x/\Gamma} \sin{(\pi k_y y)} \nonumber \\ 
    &= \sum_{k_x,k_y}
    \what{ \psi}_{k_x,k_y}(t) \phi_{k_x,k_y}(x,y), 
    \label{eq:psidecomp} 
\end{align}
where $\widehat{ \psi}_{k_x,k_y}$ is the amplitude of the ($k_x, k_y$) mode of $\psi$, and we have the constraint $\widehat{ \psi}_{k_x,k_y} = \widehat{ \psi}^*_{-k_x,k_y}$ (with the star denoting complex conjugation). We decompose $\theta$ in the same way.

\citet{Rayleigh16} has shown that the static conduction state ($\psi = \theta = 0$) bifurcates supercritically to a steady pattern of counter-rotating convection rolls vertically spanning the layer $(k_y = 1)$ when $\Ra > \Ra_c$ with
\begin{equation}
\Ra_c = \min_{k_x}{\left(\frac{\pi^4}{4\Gamma^4} \frac{(4k_x^2+\Gamma^2)^3}{k_x^2}  \right)}.
\end{equation}
The minimum is taken over integer wavenumbers $k_x$, and occurs at $k_x = 1$ provided that $\Gamma < 2^{4/3}\sqrt{1+2^{2/3}} \approx 4.05$.
Consequently, for our set-up with $\Gamma=2$, the modes $\what \psi_{1,1}$ and $\what \theta_{1,1}$ from Eq. \eqref{eq:psidecomp} become excited through a supercritical pitchfork bifurcation at $\Ra_c = 8\pi^4$. In the remainder of the article, we refer to the resulting steady pattern as the \emph{steady convection roll state} (SCRS).

A further decomposition which will aid our analysis is to partition the modes into \emph{odd modes} indicated by the letter $O$ with $k_x + k_y \in 2\mathbb{Z} + 1$, and \emph{even modes} indicated by the letter $E$ with $k_x + k_y \in 2\mathbb{Z}$ similarly to previous studies \citep{Chandra11,Verma15}.
Consistently with this notion of odd and even modes, we decompose $\psi$ as 
\begin{subequations}\label{eq:OEdecomp}\begin{align}
    \psi^O(x,y,t) &= \frac{1}{2}[\psi(x,y,t) + \psi(x+\Gamma/2,1-y,t)], \\
    \psi^E(x,y,t) &= \frac{1}{2}[\psi(x,y,t) - \psi(x+\Gamma/2,1-y,t)],
\end{align}\end{subequations}
and similarly for $\theta^O$ and $\theta^E$. Thus, $\psi^O$ and $\theta^O$ consist only of odd modes as defined above, while $\psi^E$ and $\theta^E$ consist only of even modes. Applying the decomposition \eqref{eq:OEdecomp} to equations \eqref{eq:gov}, we find that 
\begin{subequations}\label{oddeq}
\begin{align}
    \nabla^{2}\psi^O_t + \{ \psi^E, \nabla^{2}\psi^O\} + \{ \psi^O, \nabla^{2}\psi^E\} &= \Ra\Pr\theta^O_x + \Pr\nabla^4 \psi^O,\\
    \theta^O_t + \{ \psi^E, \theta^O\} + \{ \psi^O, \theta^E\}&= \psi^O_x + \nabla^2 \theta^O,
\end{align}
\end{subequations}
and
\begin{subequations}\label{eveneq}
\begin{align}
    \nabla^{2}\psi^E_t +  \{ \psi^E, \nabla^{2}\psi^E\} - \Ra\Pr\theta^E_x - \Pr\nabla^4 \psi^E &= - \{ \psi^O, \nabla^{2}\psi^O\},\\
    \theta^E_t +\{ \psi^E, \theta^E\} - \psi^E_x - \nabla^2 \theta^E &= -\{ \psi^O, \theta^O\}.
\end{align}
\end{subequations}
We observe that $( \psi^O, \theta^O )$ satisfy the linear PDEs \eqref{oddeq}, with coefficients that depend on $( \psi^E, \theta^E )$, while $( \psi^E, \theta^E )$ satisfy the autonomous non-linear PDEs \eqref{eveneq}, with forcing terms that depend on $( \psi^O, \theta^O )$. 

At the onset of steady convection the excited modes $\what\psi_{1,1}$ and $\what\theta_{1,1}$ are even modes, and indeed the SCRS has the property that $\psi^O \equiv \theta^O \equiv 0$, which remains true as we increase $\Ra$. In this case, the  system of equations \eqref{oddeq}--\eqref{eveneq} reduces to
\begin{subequations}
\label{eq:eveneqsteady}
\begin{align}
     \nabla^{2}\psi^E_t + \{ \psi^E, \nabla^{2}\psi^E\} &= \Ra\Pr\theta^E_x + \Pr\nabla^4 \psi^E,\\
     \theta^E_t + \{ \psi^E, \theta^E\} &= \psi^E_x + \nabla^2 \theta^E,
\end{align}
\end{subequations}
where the even modes are decoupled from the odd modes. The SCRS has a reflection symmetry about the vertical line which separates the counter-rotating convection rolls. Without loss of generality, we choose this vertical line to be at $x=0$. Thus, the steady state solution $(\psi^E, \theta^E)$ of the system \eqref{eq:eveneqsteady} is invariant under the symmetry
\begin{align}\label{eq:Symmetry1}
\psi^E(x,y) &\mapsto -\psi^E(-x,y), &
\theta^E(x,y) &\mapsto \theta^E(-x,y).
\end{align}
Since the zonal modes $\what \psi_{0,k_y}$ are $x$-independent, equation \eqref{eq:Symmetry1} implies that $\what\psi^E_{0,k_y}\equiv 0$. In other words, the SCRS has no zonal flow, and zonal modes can emerge only once the SCRS has become unstable.

\section{Linear stability analysis} \label{sec:Linear_stability}
In this section, we present the linear stability analysis of the SCRS. For given values of $\Pr$ and $\Ra$, we first solve numerically the steady state problem of Eq. \eqref{eq:eveneqsteady}, following the procedure described in Appendix~\ref{app:newton_method}. We then exploit the partial decoupling of odd and even modes to consider odd and even perturbations separately. We analyse odd perturbations by setting
\begin{align}
    \label{eq:pert}
    \psi(x,y,t) &= \psi^E(x,y) + \delta\mathrm{e}^{\sigma^o t} \psi^o(x,y)
\nonumber\\
&= \psi^E(x,y) + \delta\mathrm{e}^{\sigma^o t}\smashoperator[lr]{\sum_{k_x+k_y\in2\mathbb{Z}+1}} \what{\psi}^o_{k_x,k_y}\phi_{k_x,k_y}(x,y),
    \\
    \theta(x,y,t) &= \theta^E(x,y) + \delta\mathrm{e}^{\sigma^o t}\theta^o(x,y)
\nonumber\\
&=\theta^E(x,y) + \delta\mathrm{e}^{\sigma^o t}\smashoperator[lr]{\sum_{k_x+k_y\in2\mathbb{Z}+1}} \what{\theta}^o_{k_x,k_y}\phi_{k_x,k_y}(x,y),
\end{align}
where $(\psi^E,\theta^E)$ is the SCRS, $\delta \ll 1$ and 
$\psi^o$, $\theta^o$ consist only of odd modes. By substituting (\ref{eq:pert}) into equations \eqref{eq:gov} and linearising with respect to $\delta$, we obtain an eigenvalue problem for the odd growth rates $\sigma^o$ (see Appendix~\ref{app:ap_linear_stability_analysis} for details). Then, we solve this eigenvalue problem numerically and determine that SCRS is unstable with respect to odd perturbations if $\max{[\mathcal{R}(\sigma^o)]} > 0$.
An analogous approach is used to determine the corresponding even growth rates $\sigma^e$ and thus the stability of the SCRS with respect to even perturbations.

\begin{figure}
  \centerline{\includegraphics[width=0.8\textwidth]{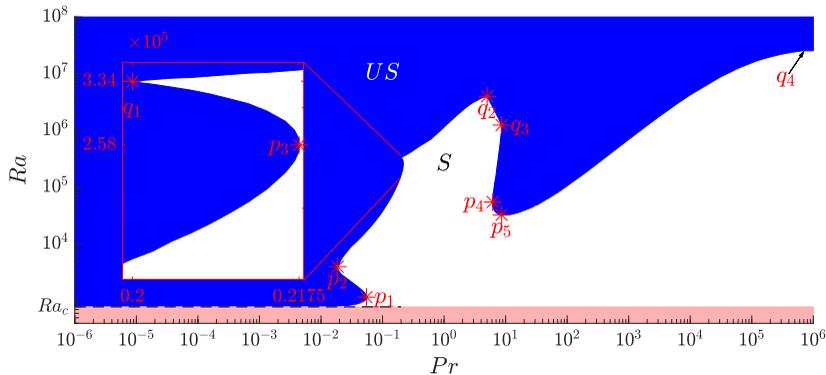}}
  \caption{The $(\Pr, \Ra)$-plane highlighting when the SCRS is linearly stable (white, $S$) and unstable (blue, $US$). Only the static conduction state is stable in the red shaded region below $\Ra = \Ra_c$. The highlighted points in red are at $p_1 = (5.48 \times 10^{-2},1.16 \times 10^3)$, $p_2 = (1.81 \times 10^{-2},4.12 \times 10^3)$, $p_3 = (0.2175,2.59 \times 10^5)$, $q_1 = (0.2,3.32 \times 10^5)$, $q_2 = (4.97,3.949 \times 10^6)$, $q_3 = (8.58,1.27 \times 10^6)$, $p_4 = (6.16,5.43 \times 10^4)$, $p_5 = (9.53, 3.24 \times 10^4)$ and $q_4 = (7 \times 10^5,2.54 \times 10^7)$.}
\label{fig:MainBlue}
\end{figure}

Figure \ref{fig:MainBlue} displays the stability of the SCRS in the $(\Pr, \Ra)$-plane, highlighting the stable ($S$) regime in white and the unstable ($US$) regime in blue.
The pink region denotes $\Ra < \Ra_c$ where the static conduction state is stable.
Although the SCRS does become unstable to even perturbations within the region $US$, we always observe that $\max{[\mathcal{R}(\sigma^o)]}>\max{[\mathcal{R}(\sigma^e)]}$ and the dominant unstable perturbation therefore consists of odd modes.
Moreover, for $\Pr\lesssim7\times10^5$ the most unstable eigenfunction has the property that $\what{\psi}^o_{k_x,k_y}$ and $\mathrm{i}\what{\theta}^o_{k_x,k_y}$ are purely real, so that $\psi^o$ and $\theta^o$ break the reflection symmetry \eqref{eq:Symmetry1}.
In particular, zonal modes $\what \psi_{0,k_y}$ with $k_y\in2\mathbb{Z}+1$ are present in the most unstable perturbation until we reach very large Prandtl numbers (see \S\ref{sec:large}).

\begin{figure}
  \centerline{\includegraphics[width=1\textwidth]{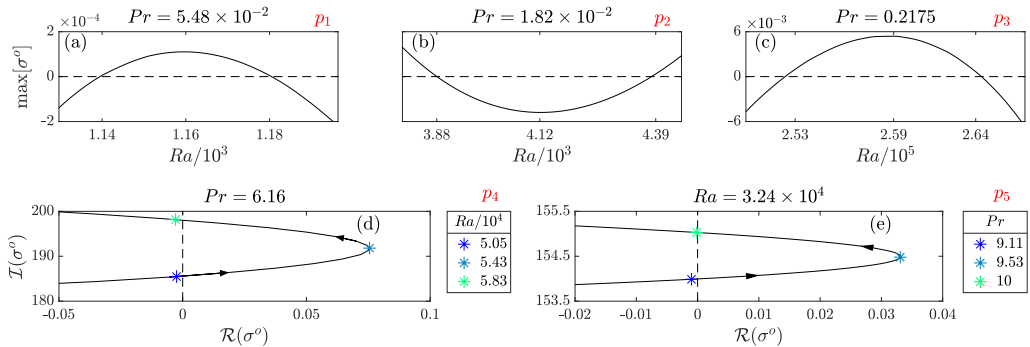}}
  \caption{The largest odd growth rate $\sigma^o$, plotted with one dimensionless parameter varied and the other fixed. The plots show the behaviour close to each of the points $p_i$ highlighted in figure \ref{fig:MainBlue}. In (a)--(c) $\sigma^o$ is real, while in (d) and (e) the motion of $\sigma^o$ in the complex plane is tracked, with the arrows indicating the direction of increasing $\Ra$ or $\Pr$.}
\label{fig:Ppoints}
\end{figure}

The points $p_i$ annotated in Figure~\ref{fig:MainBlue} highlight turning points in the stability boundary. The behaviour of $\sigma^o$ near each such point is plotted in Figure~\ref{fig:Ppoints}.
Figure~\ref{fig:Ppoints}(c) shows that, close to $\Pr=0.2$ and with increasing Rayleigh number, a real eigenvalue crosses the origin twice, and the SCRS first loses and then regains stability.
The steady state unexpectedly regains stability as the Rayleigh number increases
at point $p_1$ with $\Pr\approx0.055$. Similar behaviour has been observed by \citet{Paul12}, with $\Pr = 6.8$ and $\Gamma = 2\sqrt{2}$. At point $p_2$, the opposite occurs with the SRCS first regaining and then losing stability as $\Ra$ increases. At points $p_4$ and $p_5$ we instead find complex eigenvalues crossing the imaginary axis twice as $\Ra$ increases with fixed $\Pr$ or \textit{vice versa}.

\begin{figure}
  \centerline{\includegraphics[width=1\textwidth]{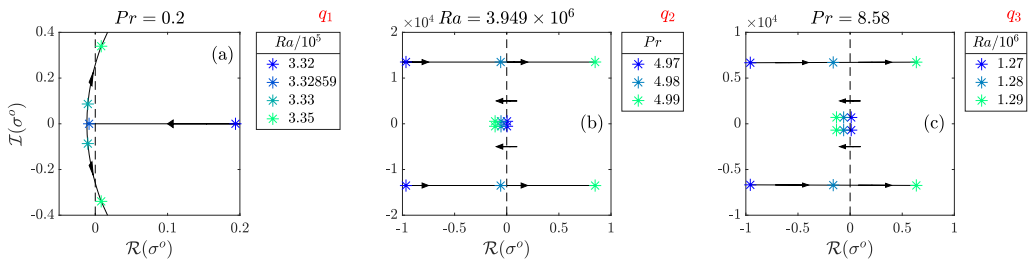}}
  \caption{The largest odd growth rate(s) $\sigma^o$, plotted with one dimensionless parameter varied and the other fixed. The plots show the behaviour close to each of the points $q_i$ highlighted in figure \ref{fig:MainBlue}. The motion of $\sigma^o$ in the complex plane is tracked, with the arrows indicating the direction of increasing $\Ra$ or $\Pr$.}
\label{fig:Qpoints}
\end{figure}

At each of the points labelled $q_i$, the stability boundary is not smooth, and there is a qualitative change in the eigenfunction to which the SCRS becomes unstable. Figure~\ref{fig:Qpoints}(a) shows how, with $\Pr=0.2$, the SCRS regains stability with a real eigenvalue crossing zero at $\Ra \approx 3.32859 \times 10^5$. With further increase in $\Ra$, the eigenvalue splits into a complex conjugate pair, which then recrosses the imaginary axis as the SCRS loses stability in a Hopf bifurcation. The cusp at $q_1$ occurs when these three events happen simultaneously.
These observations have been confirmed with direct numerical simulation (DNS) of the system \eqref{oddeq}--\eqref{eveneq}, using the pseudospectral scheme described by \cite{Winchester21}; further details are given in the Supplementary Material.
Figure~\ref{fig:Reversals} shows numerical results for the largest scale odd mode $\what{\psi}_{0,1}$ obtained with the right-hand side of \eqref{eveneq} set to $0$, so that we have persistent exponential growth or decay in the odd modes. We observe that, with increasing $\Ra$, the odd perturbations regain stability, then become oscillatory, and finally become unstable again.

\begin{figure}
  \centerline{\includegraphics[width=0.8\textwidth]{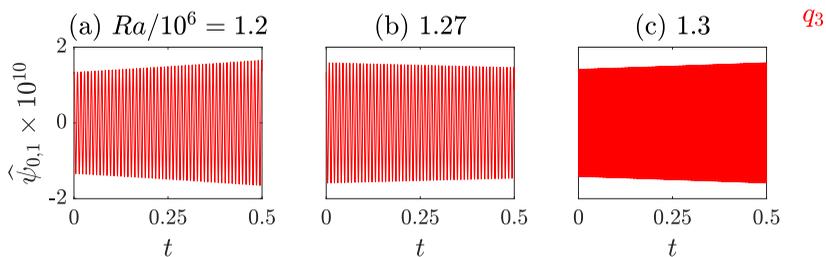}}
  \caption{Time series of the $\what \psi_{0,1}$ mode close to $q_3$ from the linearised DNS with the right-hand side of \eqref{eveneq} set to zero. We fix  $\Pr = 8.57$ and the initial conditions are the SCRS with a small perturbation. Longer time series clearly showing the exponential growth and decay are in the Supplementary Material.}
\label{fig:q3}
\end{figure}

Figure~\ref{fig:Qpoints}(c) shows that, with $\Pr=8.58$,
the SCRS regains stability through a Hopf bifurcation at $\Ra = 1.28 \times 10^6$ but loses stability shortly after at $\Ra = 1.29 \times 10^6$ with much larger oscillation frequency. The corner $q_3$ occurs when these two pairs of eigenvalues cross the imaginary axis simultaneously.
Again, this observation has been verified with DNS close to $q_3$. In Figure~\ref{fig:q3} we fix $Pr = 8.57$ and see that the odd mode $\what \psi_{0,1}$ grows at $\Ra = 1.2 \times 10^6$, decays at $\Ra = 1.27 \times 10^6$, and grows again at $\Ra = 1.3 \times 10^6$, but now with a much greater oscillation frequency. As can been seen in Figure~\ref{fig:Qpoints}(b), similar behaviour occurs at corner $q_2$. The final corner point $q_4$, which lies in the large-$\Pr$ region, will be considered in more detail in \S\ref{sec:large}. Movies showing how the dominant eigenfunction changes at each of the points $q_i$ are included in the Supplementary Material.

\begin{figure}
  \centering
  \includegraphics[width=0.8\textwidth]{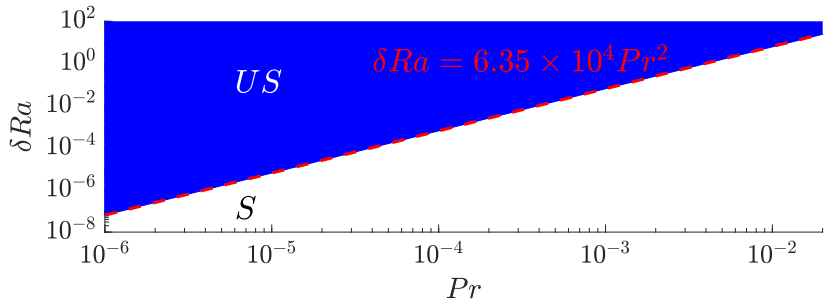}
  \caption{The $(\Pr, \delta \Ra)$-plane for small $\Pr$ highlighting when the SCRS is linearly stable (white, $S$) and unstable (blue, $US$). On the vertical axis we plot $\delta \Ra = \Ra - \Ra_c$, where $\Ra_c = 8\pi^4$. The red dashed line highlights the power-law $\delta \Ra \propto \Pr^2$, where the prefactor has been calculated by fitting.}
 \label{fig:SmallPr}
\end{figure}

Figure~\ref{fig:MainBlue} hints that the Rayleigh number at which the SCRS loses stability approaches $\Ra_c$ as $\Pr \to 0$. In Figure~\ref{fig:SmallPr} we plot $\delta \Ra:= \Ra - \Ra_c$ versus $\Pr$ and indeed see that the stability boundary follows a clear power law with $\delta \Ra \propto \Pr^2$.
When this boundary is crossed,
at small $\Pr$ the SCRS becomes unstable through a pitchfork bifurcation as a new steady state containing both even and odd modes emerges.

\begin{figure}
  \centerline{\includegraphics[width=0.8\textwidth]{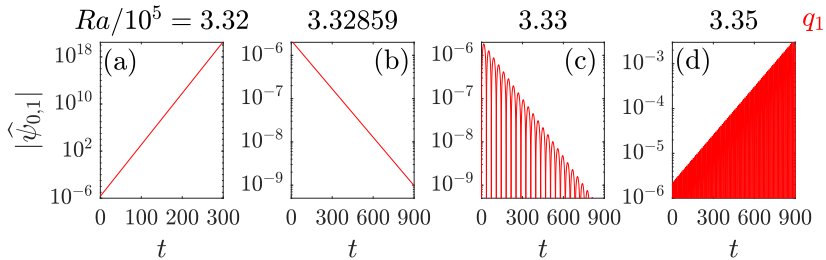}}
  \caption{Time series of the $\what \psi_{0,1}$ mode from the linearised DNS with the right-hand side of \eqref{eveneq} set to zero.  Here we fix $\Pr = 0.2$ ($q_1$) and have $\Ra = 3.32 \times 10^5$ (a), $\Ra = 3.32859 \times 10^5$ (b), $\Ra = 3.33 \times 10^5$ (c) and $\Ra = 3.35 \times 10^5$ (d). The initial conditions are the SCRS with a small perturbation.}
\label{fig:Reversals}
\end{figure}

\begin{figure}
  \centerline{\includegraphics[width=1\textwidth]{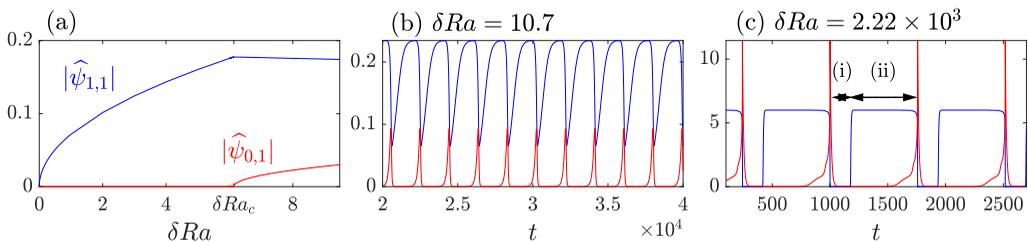}}
  \caption{(a) Bifurcation diagram plotting $\lvert \what \psi_{1,1} \rvert$ (blue) and $\lvert \what \psi_{0,1} \rvert$ (red) against $\delta \Ra$ with $\Pr = 10^{-2}$. (b) and (c) display corresponding time series of $\lvert \what \psi_{1,1} \rvert$ and $\lvert \what \psi_{0,1} \rvert$ when $\delta \Ra = 10.7$ and $\delta \Ra = 2.2 \times 10^3$, respectively.}
\label{fig:Solutions1}
\end{figure}

This prediction has been verified with fully nonlinear DNS solving Eq. \eqref{eq:gov}. using the pseudospectral scheme described in \cite{Winchester21}. 
In Figure~\ref{fig:Solutions1}(a) we plot a bifurcation diagram showing the amplitudes of the even mode $\what\psi_{1,1}$ and the odd mode $\what\psi_{0,1}$ versus $\delta\Ra$ with $\Pr=0.01$. The SCRS emerges at $\delta \Ra = 0$ and becomes unstable at $\delta \Ra = \delta \Ra_c \approx 6.135$ as the new steady state emerges. In Figure~\ref{fig:Solutions1}(b) we plot the time series of $\lvert \what \psi_{1,1} \rvert$ and $\lvert \what \psi_{0,1} \rvert$ at $\Pr=0.01$ and $\delta \Ra = 10.7$. At this point the mixed steady state has become unstable and the mode $\lvert \what \psi_{1,1} \rvert$ undergoes non-linear relaxation oscillations whilst $\lvert \what \psi_{0,1} \rvert$ exhibits bursts that quickly decay away. At $\delta \Ra = 2.22 \times 10^3$, the relaxation oscillations and the bursts persist, but are separated by a much slower buildup phase, as shown in Figure~\ref{fig:Solutions1}(c). 

In summary, the dynamics at low Prandtl number with $\delta\Ra>\delta\Ra_c$ can be partitioned into three phases, as follows.
\begin{enumerate}
    \item In the first phase we are close to the static conduction state with  $\lvert \what \psi_{1,1} \rvert,\lvert \what \psi_{0,1} \rvert \approx0$. However, when $\delta \Ra > 0$, the static conduction state is unstable, so that $\what \psi_{1,1}$ and other even modes grow exponentially. 
    \item The even modes grow until we are in the SCRS. Since $\delta \Ra > \delta \Ra_c$, the SCRS is unstable to odd perturbations, so that $\what \psi_{0,1}$ and other odd modes grow exponentially.
    \item Once $\what \psi_{1,1}$ and $\what \psi_{0,1}$ are of comparable amplitude, the system quickly collapses back to the static conduction state and the cycle continues. 
\end{enumerate}
In the following section, we carry out a weakly non-linear analysis to help us explain the observed power-law (see Figure~\ref{fig:SmallPr}) as well as the dynamics of the non-linear relaxation oscillations and of the bursts (see Figure~\ref{fig:Solutions1}).

\section{Stability analysis at small Prandtl number} \label{sec:small}
We perform a weakly non-linear analysis with the Prandtl number being our small parameter, $\epsilon = \Pr \ll 1$. The analysis is carried out for arbitrary aspect ratio $\Gamma<2^{4/3}\sqrt{1+2^{2/3}}$, and the other parameters and variables are scaled as follows:
\begin{subequations}
\begin{align}
    \Ra &= \Ra_c+r\epsilon^2= \pi^4\frac{(4+\Gamma^2)^3}{4\Gamma^4}+r\epsilon^2, \label{eq:deltaRa} \\
    t&= \frac{\tau_{_O}}{\epsilon}=
    \frac{\tau_{_E}}{\epsilon^3} , \\
    \psi^E &= \epsilon\psi_1^E(\tau_E, x,y) + \epsilon^2 \psi_2^E(\tau_E, x,y) + \mathcal{O}(\epsilon^3), \label{eq:psi^E} \\
    \theta^E &= \epsilon\theta_1^E(\tau_E, x,y) + \epsilon^2 \theta^E(\tau_E, x,y) + \mathcal{O}(\epsilon^3),\\
    \psi^O &= \epsilon^2 \psi_1^O(\tau_O, x,y) + \epsilon^3 \psi_2^O(\tau_O, x,y) + \mathcal{O}(\epsilon^4), \\
    \theta^O &= \epsilon^2 \theta_1^O(\tau_O, x,y) + \epsilon^3 \theta_2^O(\tau_O, x,y) + \mathcal{O}(\epsilon^4). \label{eq:theta^O}
\end{align}
\end{subequations}
We have introduced two time-scales $\tau_{_O}$ and $\tau_{_E}$, due to the distinct time-scale separation between the odd and even modes, observed for example in Figure~\ref{fig:Solutions1}(c). Equation \eqref{eq:deltaRa} is inspired by the power law $\delta \Ra \propto \Pr^2$. The scalings of the dependent variables are constructed such that weak nonlinearity and coupling between even and odd modes enter at the same order, as we will see below.

First we examine the evolution of the even modes. At $\mathcal{O}(\epsilon)$ and $\mathcal{O}(\epsilon^2)$ we find
\begin{subequations}
\label{eq:GovEqTwo}
\begin{align}
    \psi^E_1 &=  \frac{4+\Gamma^2}{\Gamma}\pi A \sin{(2\pi x/\Gamma)} \sin{(\pi y)}, \\
    \theta^E_1 &=  2 A \cos{(2\pi x/\Gamma)} \sin{(\pi y)},
\end{align}
\end{subequations}
and
\begin{subequations}
\begin{align}
    \psi^E_2 &= 0, \\
    \label{eq:ThetaETwo}
    \theta^E_2 &= -\pi A^2  \frac{4+\Gamma^2}{2\Gamma^2}\sin{(2 \pi y)},
\end{align}
\end{subequations}
which describe the asymptotic behaviour of the SCRS as $\delta \Ra \to 0$.

At $\mathcal{O}(\epsilon^3)$, we have
\begin{subequations}
\begin{align}
    \label{eq:eps2a}
    \nabla^{2}\psi^E_{1\tau_{_E}} + \{ \psi_1^O, \nabla^{2}\psi^O_1\}  &= \Ra_c\theta^E_{3x} + \nabla^4 \psi^E_3 + r\theta^E_{1x},\\
    \{ \psi_1^E,\theta^E_2\}&= \psi^E_{3x} + \nabla^2 \theta^E_3,
\end{align}
\end{subequations}
the solvability condition for which gives us
\begin{align}\label{eq:A}
    \frac{(4+\Gamma^2)^2}{2\Gamma^3}\pi^3\,
    \frac{\mathrm{d}A}{\mathrm{d}\tau_{E}}
    - r\frac{2\pi}{\Gamma} A +  \frac{(4+\Gamma^2)^4}{4\Gamma^7}\pi^7 A^3 =
    \frac{2\mathrm{i}}{\Gamma}\left\langle \{ \psi_1^O, \nabla^{2}\psi^O_1\}, \phi_{1,1} \right\rangle.
\end{align}
The right-hand side of equation \eqref{eq:A} is the inner product of the nonlinear forcing term with the $(1,1)$ basis function, and captures the net effect of the odd modes on the amplitude $A$ of the dominant even mode. If this term is set to zero, then equation \eqref{eq:A} reduces to the standard Landau equation governing the amplitude $A$ of weakly nonlinear perturbations to the static conduction state \citep{Fowler97}.

To evaluate the right-hand side of equation \eqref{eq:A}, we now turn to the odd modes. At lowest order we obtain the problem
\begin{subequations}
\begin{align}
    \nabla^{2}\psi^O_{1\tau_{_O}} + \{ \psi^O_1, \nabla^{2}\psi^E_1\} + \{ \psi^E_1, \nabla^{2}\psi^O_1\} &= \Ra_c\theta^O_{1x} + \nabla^4 \psi^O_1,\\
    0&= \psi^O_{1x} + \nabla^2 \theta^O_1,
\end{align}
\end{subequations}
which can be reduced to
\begin{multline}\label{eq:p1tored}
    \nabla^{2}\psi^O_{1\tau_{_O}} +
    \frac{\pi(4+\Gamma^2)}{\Gamma}A\left\{
\sin\left(\frac{2\pi x}{\Gamma}\right)\sin(\pi y),
\nabla^2\psi_1^O+\frac{\pi^2(4+\Gamma^2)}{\Gamma^2}\psi_1^O
    \right\}
    \\= \nabla^4 \psi^O_1 - \Ra_c\nabla^{-2}\psi^O_{1xx}.
\end{multline}
Using the Fourier decomposition \eqref{eq:psidecomp}, i.e.
\begin{align}
\label{eq:Bdef}
    \psi^O_1 &= \sum_{k_x+k_y \in 2\mathbb{Z}+1} \what{\psi}_{k_x,k_y}\phi_{k_x,k_y},
\end{align}
we can express equation (\ref{eq:p1tored}) as a linear dynamical system of the form
\begin{align}
    \label{eigenprob}
    \frac{\mathrm{d}\boldsymbol{\psi}^O_1}{\mathrm{d}\tau_{O}} &= (AM+D)\boldsymbol{\psi}^O_1.
\end{align} 
Here $M$ and $D$ are known constant matrices, calculated as described in Appendix~\ref{sec:appendix2}, and $\boldsymbol{\psi}^O_1=(\what{\psi}_{k_x,k_y})_{k_x+k_y\in2\mathbb{Z}+1}$ is the vector of odd modes.

We recall that $A$ evolves on the much slower time-scale $\tau_E=\epsilon^2\tau_O$.
As justified in Appendix \ref{sec:appendix2}, to leading order in $\epsilon$, it follows that $\boldsymbol{\psi}^O_1$ is proportional to the most unstable eigenvector of the problem \eqref{eigenprob}. We can therefore write
\begin{align}\label{eq:psiO1Bv}
\boldsymbol{\psi}^O_1 (\tau_O;A)\sim B(\tau_O)\boldsymbol{v}(A),
\end{align}
where
\begin{align}\label{eigenprobv}
(AM+D)\boldsymbol{v}(A)&=\sigma(A)\boldsymbol{v}(A),
\end{align} 
with $\sigma(A)$ the eigenvalue of \eqref{eigenprobv} with the largest real part, and $\boldsymbol{v}$ normalised such that $||\boldsymbol{v}||=1$. 
At leading order in $\epsilon$, the net amplitude $B$ of the odd modes thus evolves according to
\begin{align}
    \label{eq:B}
    \frac{\mathrm{d}B}{\mathrm{d}\tau_{O}} = \sigma(A)B.
\end{align}

\begin{figure}
\centering
  \includegraphics[width=0.6\textwidth]{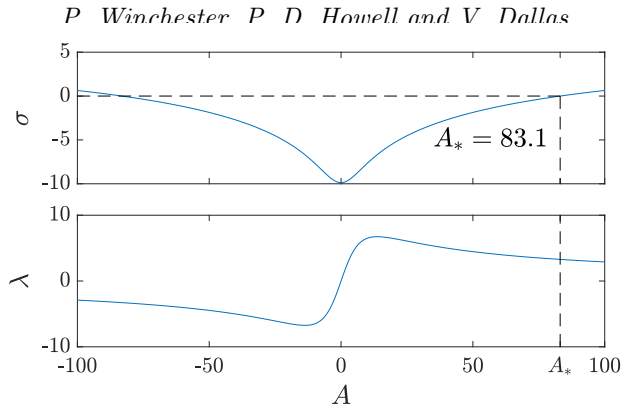}
  \caption{The functions $\sigma(A)$ and $\lambda(A)$ as they appear in \eqref{eq:system} with $\Gamma =2$. Highlighted is the point $A_*$ at which $\sigma(A_*) = 0$.}
\label{fig:SigLam}
\end{figure}

With $\psi^O_1$ decomposed as in \eqref{eq:psiO1Bv}, we can express the right-hand side of equation \eqref{eq:A} in the form
\begin{equation}
\frac{2\mathrm{i}}{\Gamma}\left\langle \{ \psi_1^O, \nabla^{2}\psi^O_1\}, \phi_{1,1} \right\rangle=-B^2\lambda(A),
\end{equation}
where
\begin{equation}\label{eq:lambdadef}
\lambda(A)=-\frac{2\mathrm{i}}{\Gamma}\left\langle\{v(A),\nabla^2v(A)\},\phi_{1,1}\right\rangle.
\end{equation}
For a given value of $\Gamma$, the functions $\sigma(A)$ and $\lambda(A)$ can both be computed once-and-for-all from equations \eqref{eigenprobv} and \eqref{eq:lambdadef}. Equations \eqref{eq:A} and \eqref{eq:B} then provide a closed two-dimensional autonomous system for the amplitudes $A$ and $B$ of the even and odd modes, respectively.

By using the scalings
\begin{align}
\label{eq:scaling}
\begin{split}
      A &= \left( \frac{4\Gamma^7}{(4+\Gamma^2)^4\pi^7} \right)^{1/3}\what{A} = a\what{A}, \\
    r = \frac{\Gamma}{2\pi a} \what{r}, \qquad \tau_E &= a \frac{(4+\Gamma^2)^2}{2\Gamma^3} \pi^3 \what{\tau_E}, \qquad B = a \frac{(4+\Gamma^2)^2}{2\Gamma^3} \pi^3 \what{B},
\end{split}
\end{align}
we normalise the system \eqref{eq:A} and \eqref{eq:B} and are left with the two ODEs (after dropping hats)
\begin{subequations}
\label{eq:system}
\begin{align}
    \label{eq:AODE}
    \dot{A} &= rA - A^3 - B^2\lambda(A), \\
     \epsilon^2 \dot{B} &= \sigma(A)B,
\end{align}
\end{subequations}
where the time derivative is taken with respect to the slow time-scale $\tau_E$. 
In Figure~\ref{fig:SigLam} we show the functions $\sigma(A)$ and $\lambda(A)$ (normalised according to \eqref{eq:scaling}) in the instance when $\Gamma = 2$. The highlighted point $A_*$ where $\sigma(A_*) = 0$ takes the value $A_* \approx 83.1$.

Since $\sigma(A)$ is an even function and $\lambda(A)$ is an odd function, we need only consider solutions of the phase plane problem (\ref{eq:system}) in the quadrant $A,B\geq0$. The critical points are:
\begin{enumerate}
    \item $(A,B) = (0,0)$, corresponding to the pure conduction state, which loses stability through the primary pitchfork bifurcation as $r$ increases through zero. This bifurcation excites the even modes in the system.
    \item\label{it:secondary} $(\sqrt{r},0)$ exists for $r > 0$ and represents the SCRS, with no odd modes. This state loses stability through a secondary pitchfork bifurcation at $r =r_c= A_*^{2} \approx 6.91 \times 10^3$ (when $\Gamma=2$), at which point the SCRS becomes unstable to odd perturbations.
    \item\label{it:stst3} $(A_*, \sqrt{(rA_* - A_*^3)/\lambda(A_*)})$ exists for $r>r_c$, when a mixed steady state including odd and even modes emerges.
    Finally, this steady state loses stability in a Hopf bifurcation at $r = r_* = A_*^2 + (2A_*^2\lambda(A_*))/(\lambda(A_*)-A_*\lambda'(A_*))$. When $\Gamma = 2$, we calculate $r_* \approx 1.53\times 10^4$.
\end{enumerate}

\begin{figure}
  \centerline{\includegraphics[width=1\textwidth]{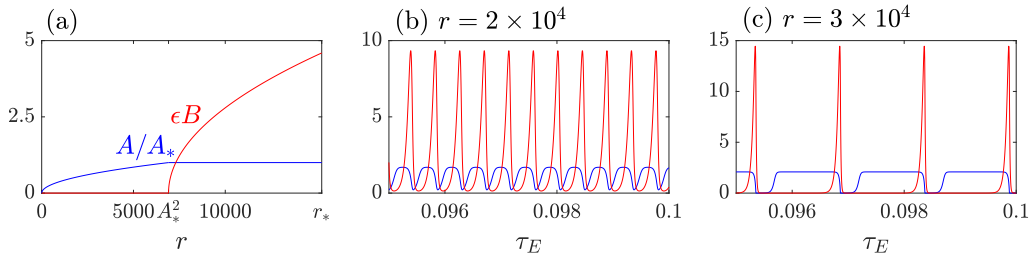}}
  \caption{(a) Bifurcation diagram plotting $A/A_*$ (blue) and $\epsilon B$ (red) against $r$ with $\Gamma=2$ and $\epsilon = 10^{-2}$. Annotated are $r=r_c= A_*^2$, the point of the secondary pitchfork bifurcation and $r= r_*$, where the Hopf bifurcation occurs. (b) and (c) display time series of $A/A_*$ and $\epsilon B$ when $r = 2 \times 10^4$ and $r = 3 \times 10^4$, respectively.}
\label{fig:BifDiagWNL}
\end{figure}

We can use the value of $r_c$ from (\ref{it:secondary}) to infer the prefactor in the power law found in \S\ref{sec:small} for the stability boundary at small $\Pr$. Rescaling back to our original variables using (\ref{eq:scaling}), we find that $r_c=\delta\Ra_c/\Pr^2 \approx 6.35 \times 10^4$ which indeed gives excellent agreement with the fit obtained in Figure~\ref{fig:SmallPr}. 

To see the bifurcations itemized above in practice, we solve equations (\ref{eq:system}) numerically. Figure~\ref{fig:BifDiagWNL}(a) is a bifurcation diagram showing how the nontrivial steady states identified in
(\ref{it:secondary}) and (\ref{it:stst3}) above emerge at $r = 0$ and $r = r_c=A_*^2$, respectively. The Hopf bifurcation occurs at $r=r_*$, beyond which point the system undergoes the nonlinear relaxation oscillations and bursts displayed in Figures~\ref{fig:BifDiagWNL}(b) and~(c) for $r = 2 \times 10^4$ and $r = 3 \times 10^4$, respectively. 

There is great qualitative agreement between the behaviours of our vastly simplified two-dimensional system and of the full system, displayed in Figures~\ref{fig:BifDiagWNL} and~\ref{fig:Solutions1}, respectively.
Although the system (\ref{eq:system}) was formally derived in the asymptotic limit as $\Pr\rightarrow0$,
similar relaxation oscillations are found in many parts of the $(\Pr, \Ra)$-plane \citep{Gol14}. The resemblance of these oscillations to the behaviour of simple predator-prey population models and the Lotka--Volterra equations has often been studied \citep{Leboeuf93,Garcia03,Garcia20,Malkov01}. In this analogy, the place of the predator population is taken by a quantity undergoing relaxation oscillations, and the place of the prey population is taken by a bursting quantity. In contrast, our model (\ref{eq:system}) has been derived systematically from the governing equations without the need for any \textit{ad hoc} closure assumptions. 

\section{Stability analysis at large Prandtl number} \label{sec:large}
\begin{figure}
  \centerline{\includegraphics[width=0.6\textwidth]{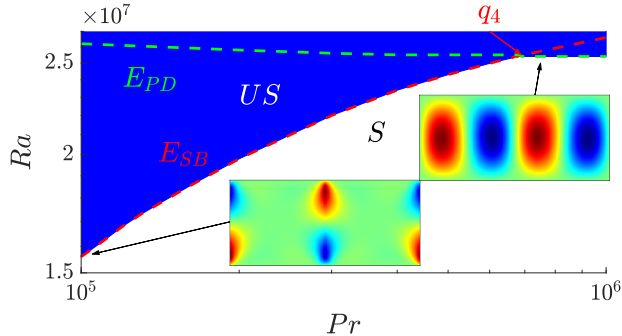}}
  \caption{The $(\Pr, \Ra)$ parameter space for large $\Pr$ highlighting when the SCRS is linearly stable (white, $S$) and unstable (blue, $US$). The point $q_4$ is where the stability boundaries corresponding to the symmetry breaking (red, $E_{SB}$) and period doubling (green, $E_{PD}$) eigenfunctions cross. The inserts show the odd eigenfunctions, $\psi^O$, associated with $E_{SB}$ and $E_{PD}$ where red, green and blue indicate $\psi^O > 0 $, $\psi^O \approx 0$ and $\psi^O < 0$, respectively. The parameters are $(\Ra,\Pr) = (1.56 \times 10^7, 10^5)$ and $(\Ra,\Pr) = (2.54 \times 10^7, 7 \times 10^5)$.}
\label{fig:LargeBlue}
\end{figure}

Figure~\ref{fig:LargeBlue} displays the $(\Pr, \Ra)$-plane for $10^5<\Pr<10^6$, highlighting where the SCRS is stable (white, $S$) or unstable (blue, $US$). 
At $q_4 = (7 \times 10^5,2.54 \times 10^7)$ there is a corner in the stability boundary and two complex conjugate pairs of eigenvalues cross the imaginary axis simultaneously as was discussed in \S\ref{sec:Linear_stability}.

The stability boundaries for the two most unstable eigenfunctions therefore cross at $q_4$, as indicated by the dashed curves in the inset in Figure~\ref{fig:LargeBlue}. As noted in \S\ref{sec:Linear_stability}, the \emph{symmetry breaking} eigenfunction (labelled $E_{SB}$) that is excited as we cross the red dashed curve has the property that $\what \psi^o_{k_x,k_y}$ and $\mathrm{i}\what \theta^o_{k_x,k_y}$ are purely real and therefore breaks the reflection symmetry between counter-rotating convection rolls in the SCRS. It follows that the horizontal velocity is an even function of $x$ which allows for instantaneous zonal flow in the solution.

The \textit{period doubling} eigenfunction (labelled $E_{PD}$) corresponding to the green dashed curve has the complementary symmetry that $\what \psi^o_{k_x,k_y}$ and $\mathrm{i}\what \theta^o_{k_x,k_y}$ are pure imaginary, and the zonal modes are therefore all zero: $\what \psi^o_{0,k_y}=0$ for all $k_y$. The horizontal velocity is now an odd function of $x$ and, although the $E_{PD}$ eigenfunction introduces odd modes to the solution, it does not break the reflection symmetry \eqref{eq:Symmetry1} in the SCRS, but instead causes a period doubling in $x$. Consequently, zonal modes at the instability switch off as we pass the point $q_4$. The qualitative change in the structure of the dominant eigenfunction is demonstrated by the inserts in Figure~\ref{fig:LargeBlue}, and movies showing how these eigenfunctions evolve in time are included in the Supplementary Material.
\begin{figure}
  \centerline{\includegraphics[width=0.6\textwidth]{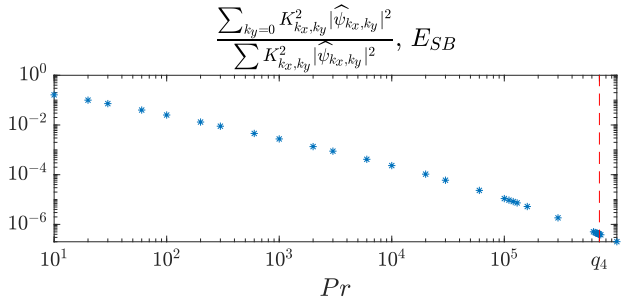}}
  \caption{The proportion of energy held in zonal modes in the eigenfunction $E_{SB}$ at the stability threshold plotted versus $\Pr$. The red dashed line highlights the point $q_4$ beyond which the eigenfunction $E_{PD}$, with no zonal modes, becomes the most unstable. We use the shorthand $K^2_{k_x,k_y} = \left(2k_{x}\pi/\Gamma \right)^2 + (k_{y}\pi)^2$.
  }
\label{fig:ZonalEnergy}
\end{figure}
As shown in Figure~\ref{fig:ZonalEnergy}, the contribution of zonal modes to the energy in the eigenfunction $E_{SB}$ at the stability boundary decreases as we increase $\Pr$. Nevertheless, $E_{SB}$ still has a small but nonzero zonal component at the point $q_4$ where the eigenfunction $E_{PD}$ takes over and zonal modes disappear completely. The instance of $E_{SB}$ in figure~\ref{fig:LargeBlue} is at the large Prandtl number $\Pr = 10^5$, and hence the zonal modes do not carry much energy. In the Supplementary Material, there is a video of $E_{SB}$ with $\Pr = 100$, in which case the presence of zonal modes is clear.

As we increase the Prandtl number past $q_4$, the green dashed line in Figure~\ref{fig:LargeBlue} appears to approach a limiting value $\Ra =  2.54 \times 10^7$, at which the SCRS is unstable for all $\Pr$.  To examine this statement, we consider the $\Pr \to \infty$ limit in which the governing equations \eqref{eq:gov} reduce to  
\begin{subequations}
\label{eq:govinf}
\begin{align}  
    \label{eq:govinfa}
    \nabla^4 \psi & =  -\Ra\theta_x,\\
    \theta_t + \{ \psi, \theta\} &= \psi_x + \nabla^2 \theta.
\end{align}
\end{subequations}
We analyse the linear stability of the SCRS for the reduced system \eqref{eq:govinf} as in \S\ref{sec:Linear_stability}, and we find that it becomes unstable at $\Ra = 2.54 \times 10^7$, consistently with the results presented in Figure~\ref{fig:LargeBlue}. It is clear from \eqref{eq:govinfa} that $\what \psi_{0,k_y} = 0$ in the $\Pr \to \infty$ limit, again in agreement with the period doubling eigenfunction identified above. We conclude that $\Pr > 7 \times 10^5$ is a large enough Prandtl number such that the system exhibits the asymptotic behaviour of identically zero zonal modes, at least linearly at the SCRS instability.

We note that the disparity between the thermal and viscous time-scales makes it very difficult to reproduce the behaviour close to $q_4$ using DNS. The linear stability analysis predicts that $E_{PD}$ oscillates with a frequency of order 10 with respect to the thermal time-scale, and it is therefore necessary to integrate the underlying equations over at least $\Pr/10\approx7\times10^4$ viscous time units to observe just one complete oscillation. At the required spatial resolution, it proved unfeasible to compute enough cycles to reliably observe the exponential growth or decay associated with the Hopf bifurcation.

\section{Conclusions} \label{sec:Conclusions}
This article concerns 2D Rayleigh--B\'enard convection with free-slip boundaries and horizontal periodicity.
The horizontal periodicity permits so-called zonal flow, with horizontal wavenumber equal to zero.
When present, the zonal flow can dominate the flow, significantly suppress convective heat transfer \citep{Gol14} and undergo random-in-time reversals in time \citep{Winchester21}.
The onset of zonal flow is therefore of great importance.

With aspect ratio $\Gamma=2$ and increasing Rayleigh number, the first state to be excited as the pure conducting state loses stability consists of steady convection rolls, referred to in the text as the SCRS. The SCRS consists of only so-called even modes and becomes unstable to odd modes as the Rayleigh number is increased.
We determine the stability boundary in the $(\Pr,\Ra)$-plane and identify corners and cusps in the boundary where qualitative changes occur to the most unstable eigenfunction. We observe three regions where the SCRS loses and subsequently regains stability as the Rayleigh number is increased.
Such behaviour seems to defy our intuition that increasing the importance of buoyancy relative to viscous and thermal dissipation should make the system less stable, although we note that reorganisation with increasing $\Ra$ has been observed experimentally by \citet{Fauve84} in convective flow of mercury at small Prandtl number. 

The excitation of odd modes is necessary, but not sufficient, for net zonal flow to exist.
The possible qualitative behaviours in our system are delineated by the points $q_1$ and $q_4$ identified in the parameter space shown in Figure~\ref{fig:MainBlue}. Below and to the left of point $q_1$, the SCRS loses stability through a pitchfork bifurcation, and the resulting steady shear flow then itself becomes unstable and undergoes nonlinear oscillations and bursts. In either case the system produces a net zonal flow.
On the other hand, to the right of point $q_4$, the dominant unstable odd eigenfunction preserves the reflection symmetry in the SCRS, and the zonal flow is identically zero.

Between points $q_1$ and $q_4$, the initial instability occurs through a Hopf bifurcation leading to oscillations in which the direction of the shear flow alternates \citep{Rucklidge96}. In this case, although the solution exhibits zonal flow instantaneously, the net (time-averaged) zonal flow is zero. This range includes the case $\Pr=30$ for which \citet{Winchester21} showed that the initial Hopf bifurcation is the first stage of a process that ultimately results in either persistent or intermittent zonal flow in the turbulent regime.
We find that zonal modes become less prominent in the most unstable odd eigenfunction as the Prandtl number increases, but are still present up to $\Pr=7\times10^5$, when they abruptly switch off at point $q_4$.

As $Pr \to 0$, the SCRS becomes unstable almost immediately as a new steady state, consisting of both even and odd modes, emerges in a pitchfork bifurcation.
We observe a power-law $\delta \Ra \propto \Pr^2$ along the stability boundary. With a weakly non-linear analysis, we derive a two-dimensional approximate model that successfully predicts the power-law in the stability boundary, including its prefactor, as well as the non-linear relaxation oscillations and bursts observed in DNS at low $\Pr$.

At large Prandtl numbers, we observe a corner in the stability boundary at $q_4 = (7 \times 10^5,2.54 \times 10^7)$ where the zonal modes in the most unstable eigenfunction switch off. As the Prandtl number increases beyond $q_4$, the stability boundary converges to the asymptotic value $\Ra = 2.54 \times 10^7$. This observation is consistent with formally taking the $\Pr \to \infty$ limit, which completely removes zonal modes. Our conclusion is that no zonal modes are excited at the SCRS instability boundary for Prandtl numbers greater than $7 \times 10^5$. Instead, an unsteady state is produced consisting of both even and odd modes. It remains open to determine when this state in turn becomes susceptible to growing zonal perturbations.

We focus here on the case of aspect ratio $\Gamma=2$.
Further analysis of the weakly nonlinear model \eqref{eq:system} suggests that the type and sequence of bifurcations that occur in the small-$\Pr$ limit may be completely different for different values of $\Gamma$. For general Prandtl number, we can anticipate that the structure of the stability boundary and the properties of the excited solutions also depend significantly on the value of $\Gamma$. In particular, if $\Gamma \geq 2^{4/3}\sqrt{1+2^{2/3}}$ then both odd and even modes become excited as $\Ra$ increases past $\Ra_c$, and it is no longer clear that the odd/even mode decomposition which proves so convenient in our set-up is still able to give some insight.

\appendix
 
\section{Computation of the SCRS}
\label{app:newton_method}
We wish to find a solution ($\psi^E$, $\theta^E$) of the steady ``even'' problem (\ref{eq:eveneqsteady}). We expand both $\psi^E$ and $\theta^E$ as in (\ref{eq:psidecomp}), i.e. in modes of the form
\begin{align}
    \phi_{k_x,k_y}(x,y) &= \mathrm{e}^{2\pi\mathrm{i} k_x x/\Gamma}\sin{( \pi k_y y)}, \qquad k_x+k_y \in 2\mathbb{Z},
\end{align}
and define the inner product 
\begin{align}
    \label{innerapend}
    \langle \phi_{k_{x_1},k_{y_1}} , \phi_{k_{x_2},k_{y_2}} \rangle = \int_0^1 \int_0^{\Gamma} \phi_{k_{x_1},k_{y_1}} \phi_{k_{x_2},k_{y_2}}^* \, \mathrm{d}x\,\mathrm{d}y = \frac{\Gamma}{2} \delta_{k_{x_1},k_{x_2}}\delta_{k_{y_1},k_{y_2}}.
\end{align}
Acting with this inner product on equations (\ref{eq:eveneqsteady}) we find, for each $(k_x,k_y)$,
\begin{subequations}\label{fg}
\begin{multline}
    \frac{\mathrm{i}}{2} \mathop{\sum_{k_{x_1} + k_{x_1} = k_x}}_{\mid k_{y_1} \pm k_{y_2} \mid = k_y} \widehat \psi_{k_{x_1}, k_{y_1}}\widehat \psi_{k_{x_2}, k_{y_2}}  K^2_{k_{x_2},k_{y_2}}G(k_{x_1},k_{x_2},k_{y_1},k_{y_2})
    \\
    +\mathrm{i}k_x\Ra\Pr\widehat \theta_{k_x, k_y} + \Pr K_{k_x,k_y}^4 \widehat \psi_{k_x, k_y}=0,
\end{multline}
\begin{multline}
    \frac{\mathrm{i}}{2} \mathop{\sum_{k_{x_1} + k_{x_1} = k_x}}_{\mid k_{y_1} \pm k_{y_2} \mid = k_y} \widehat \psi_{k_{x_1}, k_{y_1}}\widehat \theta_{k_{x_2}, k_{y_2}}  G(k_{x_1},k_{x_2},k_{y_1},k_{y_2})
    \\
    -\mathrm{i}k_x\widehat \psi_{k_x, k_y} +K_{k_x,k_y}^2 \widehat \theta_{k_x, k_y}=0,
\end{multline}
\end{subequations}
where we have introduced the shorthand
\begin{subequations}\label{eq:shorthand}
\begin{align}
K^2_{k_x,k_y} &=  \Big(\frac{2k_{x}\pi}{\Gamma} \Big)^2 + (k_{y}\pi)^2, \\
G(k_{x_1},k_{y_1},k_{x_2},k_{y_2},k_y) & =  \frac{2\pi^2}{\Gamma} \bigl(k_{x_1}k_{y_2}h(k_{y_2}, k_{y_1}, k_y)-k_{y_1}k_{x_2}h(k_{y_1}, k_{y_2}, k_y)\bigr), \\
h(x,y,z) &= \begin{cases}
    -1& \quad \text{if } x=y+z, \\
    1 & \quad \text{otherwise.} \\
  \end{cases}
\end{align}
\end{subequations}
The quadratic system of algebraic equations \eqref{fg} is solved using Newton's method, utilising the reflection symmetry \eqref{eq:Symmetry1} of the SCRS to reduce the number of unknowns.

\section{Linear stability analysis} 
\label{app:ap_linear_stability_analysis}

Here, we provide further details of the linear stability analysis discussed in \S\ref{sec:Linear_stability}.
For given $\Ra$ and $\Pr$, we construct a steady solution ($\psi^E$, $\theta^E$) to the even system (\ref{eq:eveneqsteady}) as described in Appendix~\ref{app:newton_method}.
Odd perturbations to this steady state satisfy
the ``odd'' part of the Boussinesq equations, namely the linear, time-autonomous system of PDEs (\ref{oddeq}).
We make the ansatz \eqref{eq:pert} and act on (\ref{oddeq}) with the inner product (\ref{innerapend}) to obtain an eigenvalue problem for the odd growth rate $\sigma^o$, namely
\begin{subequations}\label{oddfourier}
\begin{multline}\label{psioddstab}
     \frac{\mathrm{i}}{2} \mathop{\sum_{k_{x_o} + k_{x_E} = k_x}}_{\mid k_{y_o} \pm k_{y_E} \mid = k_y} \what \psi^o_{k_{x_o},k_{y_o}}\what\psi^E_{k_{x_E},k_{y_E}}
    G(k_{x_E},k_{y_E},k_{x_o},k_{y_o},k_y)\left[K^2_{k_{x_o},k_{y_o}} - K^2_{k_{x_E},k_{y_E}}\right]
    \\
    -\mathrm{i}k_x\Ra\Pr\what\theta^o_{k_x,k_y} - \Pr K_{k_x,k_y}^4\what\psi^o_{k_x,k_y}
    =\sigma^o K_{k_x,k_y}^2 \what \psi^o_{k_x,k_y},
    \end{multline}
\begin{multline}\label{thetaoddstab}
     \frac{\mathrm{i}}{2} \mathop{\sum_{k_{x_o} + k_{x_E} = k_x}}_{\mid k_{y_o} \pm k_{y_E} \mid = k_y} \left[ \what\theta^o_{k_{x_o},k_{y_o}}\what\psi^E_{k_{x_E},k_{y_E}} - \what\psi^o_{k_{x_o},k_{y_o}}\what\theta^E_{k_{x_E},k_{y_E}}\right]G(k_{x_E},k_{y_E},k_{x_o},k_{y_o},k_y)
\\
    + \mathrm{i}k_x \what\psi^o_{k_x,k_y} - K_{k_x,k_y}^2\what\theta^o_{k_x,k_y}
    =\sigma^o \what \theta^o_{k_x,k_y}.
\end{multline}
\end{subequations}
where $K^2$, $G$ and $h$ are as in (\ref{eq:shorthand}).

Similarly, for even perturbations, we let $\psi = \psi^E + \delta \psi^e e^{\sigma^e t}$ and $\theta = \theta^E + \delta \theta^e e^{\sigma^e t}$, and linearise (\ref{eveneq}) with respect to $\delta$ to obtain the same system as in (\ref{oddfourier}), but with $o \mapsto e$.

\section{Weakly non-linear analysis} \label{sec:appendix2}

In this appendix, we provide some details on the construction of the eigenvalue problem (\ref{eigenprob}) and the derivation of the amplitude equation (\ref{eq:B}) for the odd modes.

First, substitution of the decomposition \eqref{eq:Bdef} into the first-order odd equation \eqref{eq:p1tored} results in the linear system
\begin{multline}\label{appC1}
      \frac{\mathrm{d}\what{\psi}_{k_x,k_y}}{\mathrm{d}\tau_O} 
      =
      A(\tau_E)
      \sum_{(n,m)\in S} \frac{\pi\Gamma}{2}\left(K^2_{n,m} - \frac{K^4_{1,1}}{K_{n,m}^2}\right)\bigl(nh(n,k_x,1)-mh(m,k_y,1)\bigr)\what{\psi}_{n,m}
      \\
    +\frac{\Gamma}{2}\left( \frac{4\pi^2\Ra_c
    k_x^2}{\Gamma^2K^{4}_{k_x,k_y}}-K^{2}_{k_x,k_y}  \right)\what{\psi}_{k_x,k_y},  
\end{multline} 
where  $S$ denotes the set
\begin{align}
     S &= \bigl\{(k_x+1,k_y+1), (k_x-1,k_y+1), (k_x-1,k_y+1), (k_x-1,k_y-1)\bigr\},
\end{align}
and $K^2_{k_x,k_y}$ and $h$ are as in (\ref{eq:shorthand}).
The first and second terms on the right-hand side of equation \eqref{appC1} respectively define the elements of the matrix $M$ and of the diagonal matrix $D$ in equation \eqref{eigenprob}, which takes the form
\begin{align}\label{eigenprobslow}
    \epsilon^2 \frac{\mathrm{d}\boldsymbol{\psi}^O_1}
    {\mathrm{d}\tau_E} &= \bigl(A(\tau_E)M+D\bigr)\boldsymbol{\psi}^O_1
\end{align}
on the slow time-scale $\tau_E$.

We seek the solution for $\boldsymbol{\psi}^O_1$ as a WKB asymptotic expansion of the form
\begin{align}
    \boldsymbol{\psi}^O_1(\tau_E) \sim \mathrm{e}^{\varphi(\tau_E)/\epsilon^2}\boldsymbol{C}(\tau_E;\epsilon) \sim \mathrm{e}^{\varphi(\tau_E)/\epsilon^2}\sum_{n} \boldsymbol{C}_n(\tau_E)\epsilon^{2n},
\end{align}
following which \eqref{eigenprobslow} becomes
\begin{align}\label{eigenprobC}
    \left(\dot{\varphi} I -A(\tau_E) M - D \right)\boldsymbol{C} + \epsilon^2\dot{\boldsymbol{C}} = \boldsymbol{0}.
\end{align}
To leading order in $\epsilon$, we find that $\dot{\varphi}=\sigma(A)$ and $\boldsymbol{C}_0=b(\tau_E)\boldsymbol{v}(A)$, where $\sigma$ and $\boldsymbol{v}$ satisfy the eigenvalue problem
\begin{align}
    \left(\sigma I -A M - D \right)\boldsymbol{v} = \boldsymbol{0}.
\end{align}

For a modal truncation such that our system is of size $N$ we (in general) have $N$ eigenvalues $\sigma_{j}(A)$, ordered such that $\mathcal{R}(\sigma_{j}(A)) \geq \mathcal{R}(\sigma_{j+1}(A))$, and corresponding eigenvectors $\boldsymbol{v}_j(A)$, assumed to be normalised such that $\|\boldsymbol{v}_j(A)\|_{l^2} = 1$.
The general leading order solution for $\boldsymbol{\psi}^O_1$ is then
\begin{align}\label{psi1olo}
    \boldsymbol{\psi}^O_1(\tau_E) \sim \sum_{j=1}^N  b_j(\tau_E) \mathrm{e}^{\varphi_j(\tau_E)/\epsilon^2}\boldsymbol{v}_j\bigl(A(\tau_E)\bigr),
\end{align}
with $\dot{\varphi}_j(\tau_E) = \sigma_j(A(\tau_E))$.

Since $\mathcal{R}(\sigma_{j})\geq\mathcal{R}(\sigma_{j+1})$, the solution \eqref{psi1olo} is dominated by the first term in the sum. In fact, we observe that the dominant eigenvalue $\sigma_1$ is real and that $\mathcal{R}(\sigma_{2}) < 0$, meaning that all terms except the first are exponentially small after a short transient.
We thus have
\begin{align}
    \boldsymbol{\psi}^O_1(\tau_E) \sim   B(\tau_E) \boldsymbol{v}_1\bigl(A(\tau_E)\bigr),
\end{align}
where the leading-order amplitude is given by $B(\tau_E) = b_1(\tau_E) e^{\varphi_1(\tau_E)/\epsilon^2}$.
Taking derivatives with respect to $\tau_E$ leaves us with 
\begin{align}
    \epsilon^2 \dot{B}(\tau_E) \sim \sigma_1\bigl(A(\tau_E)\bigr)B + O(\epsilon^2),
\end{align}
which is equivalent to equation \eqref{eq:B}.
An equation for $b_1$ can be obtained in principle from the solvability condition for $\boldsymbol{C}_1$ in \eqref{eigenprobC}, but does not affect the leading-order evolution of $B$.

\bibliographystyle{jfm}
\bibliography{jfm-instructions}

\end{document}